\documentclass[11pt]{article}
\usepackage{graphicx}
\usepackage{epstopdf}

\begin{document}



\def\xslash#1{{\rlap{$#1$}/}}
\def \p {\partial}
\def \dd {\psi_{u\bar dg}}
\def \ddp {\psi_{u\bar dgg}}
\def \pq {\psi_{u\bar d\bar uu}}
\def \jpsi {J/\psi}
\def \psip {\psi^\prime}
\def \to {\rightarrow}
\def\bfsig{\mbox{\boldmath$\sigma$}}
\def\DT{\mbox{\boldmath$\Delta_T $}}
\def\xit{\mbox{\boldmath$\xi_\perp $}}
\def \jpsi {J/\psi}
\def\bfej{\mbox{\boldmath$\varepsilon$}}
\def \t {\tilde}
\def\epn {\varepsilon}
\def \up {\uparrow}
\def \dn {\downarrow}
\def \da {\dagger}
\def \pn3 {\phi_{u\bar d g}}

\def \p4n {\phi_{u\bar d gg}}

\def \bx {\bar x}
\def \by {\bar y}

\begin{center} 
{\Large\bf   Fracture Functions in Different Kinematic Regions and Their Factorizations   }
\par\vskip20pt
X.P.  Chai$^{1,2}$,  K.B. Chen $^{1,2}$, J.P. Ma$^{1,2,3}$ and X.B. Tong$^{1,2}$    \\
{\small {\it
$^1$ Institute of Theoretical Physics, Chinese Academy of Sciences,
P.O. Box 2735,
Beijing 100190, China\\
$^2$ School of Physical Sciences, University of Chinese Academy of Sciences, Beijing 100049, China\\
$^3$ School of Physics and Center for High-Energy Physics, Peking University, Beijing 100871, China
}} \\
\end{center}
\vskip 1cm
\begin{abstract}
Fracture functions are parton distributions of an initial hadron in the presence of an almost collinear particle observed in the final state.
They are important ingredients in QCD factorization for processes where a particle  is produced diffractively.  
There are different fracture functions for a process in different kinematic regions.  We take the production of a lepton pair combined with a diffractively produced particle in hadron collisions to discuss this.  Those fracture functions can be factorized further if there are large energy scales involved.  We perform one-loop calculations 
to illustrate the factorization in the case with the diffractively produced particle as a real photon. 
Evolution equations of different fracture functions are derived from our explicit calculations. They agree 
with expectations. These equations can be used for re-summations of large log terms in perturbative expansions. 
      
\vskip 5mm
\noindent
\end{abstract}
\vskip 1cm

\noindent
{\bf 1. Introduction} 

It is well-known that QCD factorization can be used to predict inclusive productions of a particle with large transverse momentum in hadron-collisions or in Semi-Inclusive DIS(SIDIS).  The predictions from QCD factorizations are made in terms of parton distribution functions and parton fragmentation functions.  But, if the transverse momentum $k_\perp$ of the produced particle is very small 
in hadron collisions or the particle produced in SIDIS is in target fragmentation region, the predictions
with parton distribution functions and parton fragmentation functions fail, because perturbative coefficient 
functions in the factorization become divergent as powers of $\ln k_\perp$.     

\par 
It has been proposed in \cite{TrVe} to use fracture functions to describe particle production in target fragmentation region of DIS. Experimentally,  evidences of such a production have been found at HERA\cite{HERA}.  An one-loop calculation in \cite{DGSIDIS}  of SIDIS shows that QCD factorization with fracture functions can be made  in the target fragmentation region in the sense that perturbatively calculable part 
is finite.  At the moment, most information about fracture functions comes from analysis of HERA data, e.g., in \cite{HERADF1, HERADF2}.  For hadron collisions,  factorizations with fracture functions for the production of one particle combined with
a lepton-pair have been shown to hold at one-loop level in \cite{FCLT, FC}, where the produced particle 
is in the forward- or backward regions.  We notice that fracture functions used in these works are for the case 
where the transverse momentum of the particle in the final state is not observed or integrated over.  

\par 
Fracture functions, also called as diffractive parton distributions,  are parton distributions of an initial hadron with one diffractively produced  particle observed in the final state.  There are different fracture functions, which can be used in different kinematic regions of a process.     
Taking the production of a lepton pair with large invariant mass in hadron collisions associated with a diffractively produced particle as 
an example,  there are two kinematic regions where predictions can be made with fracture functions. One is the region in which the lepton pair is with large transverse momentum.  The differential cross-section in this region can be factorized with the integrated fracture functions of partons, whose transverse momenta are integrated over.  Another region is specified by the small transverse momentum of the lepton pair. In this case, 
the transverse momenta of partons can not be neglected, because the small transverse momentum of the lepton pair comes partly from that of partons.  For this region, one needs to introduce 
Transverse Momentum Dependent(TMD) fracture functions.  
Integrated fracture functions have been defined in \cite{BeSo}, where their asymptotic behavior 
has been derived for the region where the momentum fraction of the struck parton approaches 
its maximal value.  In \cite{ABAK} TMD quark fracture functions have been classified for a polarized 
spin $1/2$ hadron in the initial state.   
\par 
In this work, we discuss the factorizations for the production of lepton pair associated with a diffractively produced photon in hadron collisions in the two different kinematic regions discussed in the above. 
These factorizations can be proved following the proof for Drell-Yan processes in \cite{CSS}.   
Then we study the factorization properties of integrated- and TMD quark fracture functions
in the case that the produced particle is a real photon.  We show that at one-loop level the integrated fracture 
function can be factorized with standard twist-2 parton distributions and fragmentation functions if the transverse momentum of the photon is large.  We also show that the TMD quark fracture function 
in the impact parameter $b$ space can be factorized similarly, or with the integrated fracture function, 
if the parameter $b$ is small.    
Although we take the case of the photon, the general features of factorizations of fracture functions 
are the same as those in the case of a hadron. Through our results we can derive 
evolution equations and Collins-Soper equation of TMD fracture function.  With arguments these equations are expected 
to be the same as those of twist-2- or TMD parton distributions.  In this work we give their explicit derivations. Using these equations one can perform re-summations of large log's in perturbative coefficient functions in collinear factorizations.  

\par 
In this work we are concentrated on the case where a photon is produced diffractively.  If we consider the production of a hadron instead of the photon,  the factorization of differential cross-sections  with fracture functions can be failed for hadron collisions as shown in \cite{CFS}. For inclusive production 
of a single hadron in target fragmentation region of SIDIS,  one can prove  the factorization with fracture functions, as discussed in \cite{BeSo, DIFCOL}.  The results about  factorizations of fracture functions 
in a hadron are more complicated than those presented in this work. They are under preparing.

\par 
Our paper is organized as in the following:  In Sect.2 we give and discuss factorization formulas with fracture functions for differential cross-section of the production of a lepton pair associated with a diffractively produced photon from hadron 
collisions in two kinematical regions. We also discuss the equivalence between TMD factorizations 
with differently defined TMD parton distributions.  In Sect.3 we show at one-loop level that the integrated fracture function at large transverse momentum is factorized with parton distribution functions and fragmentation functions. 
A part of the evolution equation is the same as that of quark distribution as shown from our explicit result. In Sect.4 we study the TMD quark fracture function at large transverse momenta or at small impact parameter $b$. We show that at one-loop level the function can be factorized in terms of parton distribution functions and fragmentation functions 
if the transverse momentum of the final photon is large.  We also show that in the case of the small transverse momentum 
the TMD fracture function in impact $b$-space is factorized with the integrated fracture function.  
From our results we derive the evolution equation 
of the renormalization scale $\mu$ and Collins-Soper equation. These equations are the same 
as those of TMD quark distribution as expected. 
Sect.5 is our summary.

\par\vskip20pt
\noindent
{\bf 2. Factorizations of Differential Cross-sections} 

\par\vskip10pt
\noindent
{\bf 2.1. Notations} 
\par 
We will use the  light-cone coordinate system, in which a
vector $a^\mu$ is expressed as $a^\mu = (a^+, a^-, \vec a_\perp) =
((a^0+a^3)/\sqrt{2}, (a^0-a^3)/\sqrt{2}, a^1, a^2)$ and $a_\perp^2
=(a^1)^2+(a^2)^2$. We introduce two light cone vectors $n^\mu = (0,1,0,0)$ and  $l^\mu =(1,0,0,0)$. The transverse metric is given by $g_\perp^{\mu\nu} = g^{\mu\nu} - n^\mu l^\nu - n^\nu l^\mu$. 
\par 

We consider the process: 
\begin{equation} 
h_A(P_A)  + h_B(P_B)  \to \gamma^* (q)  +  \gamma (k) + X, 
\label{DYP}  
\end{equation}
where the virtual photon will decay into an observed lepton-pair.  The momenta are indicated in brackets.   
We take a frame in which the momenta in the process are given by: 
\begin{eqnarray}
  P_A^\mu = (P_A^+, 0,0,0), \quad P_B^\mu =(0,P_B^-,0,0), \quad q^\mu =(q^+,q^-, \vec q_\perp), \quad 
    k^\mu = (k^+, k^-, \vec k_\perp), \quad q^2 = Q^2.  
\end{eqnarray} 
We are interested in the kinematical region:
\begin{equation} 
   Q^2 \gg \Lambda_{QCD}^2, \quad k^\mu \sim Q (1, \lambda^2, \lambda,\lambda),  \quad \lambda\ll 1,  
\end{equation}
i.e.,  the observed photon is produced diffractively.  We define the hadronic tensor for the process in Eq.(\ref{DYP})  as: 
\begin{equation}
W^{\mu\nu}  = \sum_X \int \frac{d^4 x}{(2\pi)^4} e^{iq \cdot x} \langle h_A(P_A), h_B(P_B)   \vert
    \bar q(0) \gamma^\nu q(0) \vert X, \gamma (k) \rangle \langle \gamma (k),X \vert \bar q(x) \gamma^\mu q(x) \vert
      h_A(P_A), h_B(P_B)\rangle, 
\end{equation}
where the spin of the initial state is averaged and the spin of the final photon is summed.  
For simplicity we set the charge fraction $e_q$ of quarks as $1$.   
We consider the differential cross-section, in which only the momenta of the photon and the lepton pair 
are observed. It is given by:
\begin{eqnarray}
(2\pi)^3 2 k^0 \frac{ d\sigma} {d^4 q d^3 k } = \frac{e^4 }{12\pi s Q^2} W_{\mu\nu} \biggr ( \frac{q^\mu q^\nu}{q^2} - g^{\mu\nu} \biggr ) 
\end{eqnarray}  
with $s= 2P_A^+ P_B^-$.  

\par 
With $Q^2 \gg \Lambda_{QCD}^2$,   there are effects which are calculable with perturbative QCD. 
One can separate or factorize nonperturbative effects from perturbative effects.  
The hadronic tensor takes different factorization forms in different regions of the transverse momentum $q_\perp$ of the lepton pair.  

\par\vskip10pt

\par\vskip5pt
\begin{figure}[hbt]
	\begin{center}
		\includegraphics[width=6cm]{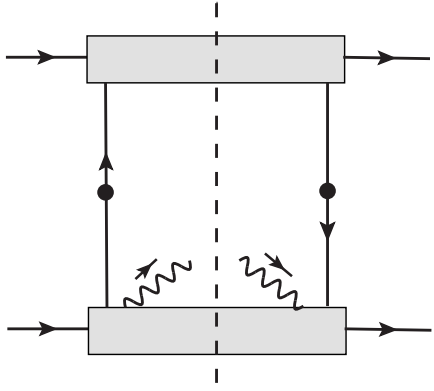}
	\end{center}
	\caption{Tree-level diagram for the hadronic tensor. }
	\label{TreeL}
\end{figure}

\noindent 
{\bf 2.2. Factorization with TMD fracture function} 
\par 
We discuss here the case with $q_\perp \sim Q\lambda$.  At tree-level  $W^{\mu\nu}$ receives the contribution from Fig.1, 
where a quark comes from hadron $h_A$ annihilates with an antiquark $\bar q$ from $h_B$ into the virtual photon. Since 
the real photon has $k_\perp\sim Q\lambda$ with $\lambda\to 0$,  its production is nonperturbative.  The description of 
the production is given by the fracture function indicated by the lower bubble in Fig.1.  Since $q_\perp$ is small, 
one can not neglect the transverse momenta of partons.     

\par 

Following general arguments for the factorization of Drell-Yan processes and SIDIS in \cite{CSS, JMY}, the hadronic tensor can be factorized with TMD fracture function of $h_A$ and TMD parton distribution of $h_B$. 
We define the TMD fracture function of $h_A$  in impact parameter $b$-space with the transverse vector $b^\mu =(b^1, b^2)$
as: 
\begin{eqnarray} 
  {\mathcal  F}_{q/h_A}  (x, b ,  \xi, k_\perp) =  \int \frac{d \lambda  }{4\pi } e^{  i  x P_A^+ \lambda  } 
     \sum_X \langle h_A (P_A) \vert \bar q (0) {\mathcal L}_u (0) \gamma^+  \vert X,  \gamma (k) \rangle 
  \langle \gamma (k),  X \vert 
     {\mathcal L}^\dagger _u (\lambda n + b)  q (\lambda n +b ) \vert h_A (P_A) \rangle. 
\label{DEFFFb}           
\end{eqnarray}
Transforming it into the momentum space the function in the transverse momentum space is obtained:   
\begin{eqnarray} 
  {\mathcal F}_{q/h_A}  (x, k_{A\perp},  \xi, k_\perp) =   \int \frac{d^2 b }{(2\pi)^2 } e^{i   k_{A\perp}  \cdot b  } 
   {\mathcal F}_{q/h_A}  (x, b ,  \xi,k_\perp), 
\label{DEFFF}           
\end{eqnarray} 
with  $k^+ = \xi P_A^+$.  Here, the quark as the parton 
carries not only the part of the hadron momentum given by $k_A^+ = x P_A^+$,  but also a transverse momentum $k_{A\perp}$. The momentum 
fraction $x$ of the parton  is in the region $1-\xi > x >0$.  ${\mathcal L}_u$ 
is a gauge link to make the definition gauge invariant.    
Because  $k_{A\perp}$ is not integrated over,   light-cone singularities will appear if one defines TMD parton distributions with a gauge link 
along light-cone directions. Light-cone singularities are also called as rapidity divergences.   
We regularize the singularities as in \cite{CSS, JMY} by introducing 
the gauge link slightly off light-cone direction: 
\begin{eqnarray}
{\mathcal L}_u (\xi) = P \exp \left ( -i g_s \int_{-\infty}^0  d\lambda
     u\cdot G (\lambda u + \xi) \right ) ,
\end{eqnarray} 
with $u^\mu=(u^+,u^-,0,0)$ and $u^-\gg u^+$. With the small- but finite $u^+$ light-cone singularities are regularized. 
The physical interpretation of the defined TMD fracture function is the parton distribution 
of $h_A$ in the presence of  an observed photon in the final state.  In the definition of Eq.(\ref{DEFFFb}) one should also add the 
gauge link 
of  electromagnetic field to make the definition $U_{em}(1)$ gauge invariant. In this work 
we take the light-cone gauge $n\cdot A =0$ for QED,  the electromagnetic gauge links are $1$ and can be omitted. We work only with the leading order of QED.

\par
The TMD parton distribution of $h_B$ is defined similarly to Eqs.(\ref{DEFFFb},\ref{DEFFF}) as:
\begin{eqnarray}
  f_{\bar q/h_B}  (y,  b ) &=&  - \int \frac{d \lambda  }{4\pi  } e^{- i  y P_B^- \lambda }   
      \langle  h_B (P_B ) \vert \bar q (0) \gamma^-  {\mathcal L}_v (0) {\mathcal L}_v^\dagger  (\lambda l +b )     q ( \lambda l +b  ) \vert h_B (P_B) \rangle , 
\nonumber\\ 
    f_{\bar q/h_B }  (y, k_{B\perp} )  &=&  \int \frac {d^2 b }{ (2\pi)^2 } e^{ - i   k_{B\perp}\cdot b    }  f_{\bar q/h_B}  (x,  b ), 
\label{DEFhb}            
\end{eqnarray} 
where  the anti-quark as the parton carries the momentum $\tilde k_B^\mu = (0, y P_B^-, k_{B\perp}^1,k^2_{B\perp})$.  The gauge link ${\mathcal L}_v$  here is along the direction $v^\mu = (v^+,v^-,0,0)$ and $v^+ \gg v^-$.  The gauge links along non-light-like directions introduce 
the dependence of TMD parton distribution and TMD fracture function on hadron's energies or    
the parameters 
\begin{equation} 
  \zeta_v^2 = \frac{ (2 y v\cdot P_B)^2}{ v^2},\quad \zeta_u^2 = \frac{  (2 x u\cdot P_A)^2} {u^2}  
\end{equation} 
respectively. The evolution of TMD fracture function and parton distribution along these parameters are governed 
by Collins-Soper equations, which are useful for resummations of large log's of $k_\perp/Q$ or $q_\perp/Q$   in perturbative 
coefficient functions of collinear factorization.   

\par 
Because the transverse momentum $q_\perp$ can also be generated from radiations of soft gluons, one 
needs not only TMD fracture function and TMD parton distribution for the factorization, but also 
a soft factor. The needed soft factor  for the subtraction of soft gluons is defined as:
\begin{equation}
      S(b, \rho) =  \frac{1} {N_c}
      \langle 0 \vert {\rm Tr}  \left [ {\mathcal L}^\dagger_v ( b) {\mathcal L}_u (b)
      {\mathcal L}_u^\dagger (0) {\mathcal L}_v (0) \right ]  \vert 0\rangle , 
 \label{SF1}      
 \end{equation} 
with $\rho^2 = (2 u\cdot v)^2/(u^2 v^2)$.  
At tree-level one has $S^{(0)} (b, \rho ) = 1$.  Following the studies in \cite{CSS, JMY}, 
the hadronic tensor in the kinematical region $q_\perp \sim Q\lambda$ can be factorized as:  
\begin{eqnarray}
W^{\mu\nu} &=&  - g_\perp^{\mu\nu} \frac{1}{N_c} H (Q,\zeta_u,\zeta_v) \int d^2 b \  e^{-i b\cdot q_\perp} 
  f_{\bar q/ h_B} (y, b, \zeta_v) {\mathcal F}_{q/h_A}  (x, b, 
     \xi, k_\perp, \zeta_u) S^{-1}(b, \rho)  +\cdots , 
\label{TMDFCF}      
\end{eqnarray} 
with $q^\mu =(x P_A^+, y P_B^-, q_\perp^1,q_\perp^2)$. There are power corrections 
to the factorization formula denoted by $\cdots$.  They are suppressed by powers of $\lambda$ or $\Lambda_{QCD}/Q$.  
$H$ is the perturbative coefficient.  It is the same as that in TMD factorization of Drell-Yan process. From  the explicit calculation in \cite{JMYDY,GZMa}, the perturbative coefficient is:  
\begin{eqnarray} 
 H  &=& 1 + \frac{\alpha_s C_F}{2\pi}
  \left [ 2\pi^2 -4
    -\ln\frac{\mu^2}{Q^2} \left ( 1+ \ln\rho^2  \right )
     -\ln\rho^2 +\frac{1}{2}\left ( \ln^2\frac{Q^2}{\zeta_v^2}
      +\ln^2\frac{Q^2}{ \zeta_u^2} \right ) \right ] +{\mathcal O}(\alpha_s^2). 
\end{eqnarray}
 
\par 
In Eq.(\ref{TMDFCF}) we only give the contribution from the quark fracture function of $h_A$ and the TMD antiquark distribution of $h_B$. There are contributions from the TMD quark distribution of $h_B$ and the TMD antiquark fracture function  of $h_A$. They can be obtained through transformation of charge-conjugation,  i.e.,  by reversing the directions 
of quark lines in Fig.1. 
\par 
In the above, the used TMD fracture function, parton distribution function are called as unsubtracted ones. 
They are defined with non light-cone gauge links for regularizing light-cone singularities.  
It is noted that the TMD fracture function here is defined with the same operators used to define TMD parton distributions, if one ignores the photon in the intermediate state in Eq.(\ref{DEFFFb}). 
There are different methods to regularize or eliminate the singularities. With different methods TMD parton distributions can be defined differently.  In \cite{JC1} light-cone gauge links are used to define unsubtracted TMD 
quark distributions. The light-cone singularities are cancelled by a soft factor which is different than the one here.  In the framework of Soft Collinear 
Effective Theory(SCET),  only light-cone gauge links are used. The light-cone singularities are regularized 
by the so-called $\delta$-regulators in \cite{EIC} or by the $\nu$-regulators in \cite{CJNR}.  Although 
TMD factorizations can be different with different TMD parton distributions, they are equivalent to each other. In the below we discuss the equivalence with Drell-Yan processes as an example, i.e., the process 
in Eq.(\ref{DYP}) without the real photon in the final state.  

\par 
The unsubtracted TMD quark distribution of $h_A$ with non light-cone gauge links is similar to Eq.(\ref{DEFFFb}):
\begin{eqnarray} 
   f_{q/h_A}  (x, b, \zeta_u) =  \int \frac{d \lambda  }{4\pi } e^{  i  x P_A^+ \lambda  } 
       \langle h_A (P_A) \vert \bar q (0) {\mathcal L}_u (0)   \gamma^+
     {\mathcal L}^\dagger _u (\lambda n + b)  q (\lambda n +b ) \vert h_A (P_A) \rangle, 
\label{DEFfq}           
\end{eqnarray}  
and the unsubtracted antiquark distribution $f_{\bar q/h_B}$ of $h_B$ is defined in Eq.(\ref{DEFhb}). 
It depends on the parameter $\zeta_v$. With the defined soft factor in Eq.(\ref{SF1}) we can define 
the subtracted TMD parton distribution as: 
\begin{equation} 
   q (x, b, \zeta_u, \rho) = \frac{f_{q/h_A}  (x,b, \zeta_u)}{\sqrt{ S(b, \rho)}}, \quad 
    \bar q(x, b, \zeta_v, \rho) = \frac{f_{\bar q/h_B }  (x,b, \zeta_v)}{\sqrt{ S(b, \rho)}} . 
\end{equation}     
With these subtracted distributions, the hadronic tensor of Drell-Yan processes in the region 
of small $q_\perp$ is factorized as\cite{JMY}:
\begin{eqnarray}
W^{\mu\nu} =  - g_\perp^{\mu\nu} \frac{1}{N_c} H (Q,\zeta_u,\zeta_v) \int d^2 b \  e^{-i b\cdot q_\perp} 
   q (x, b, \zeta_u, \rho) \bar q (y, b, \zeta_v, \rho)  +\cdots , 
\label{TMDDY}      
\end{eqnarray} 
with the same perturbative coefficient $H$ as given before.  It should be noted that here the light-cone singularities in the TMD quark- and TMD anti-quark distributions are regularized independently, i.e., 
$\zeta_u$ and $\zeta_v$ are independent parameters.  The parameter $\rho$ from the soft factor 
is fixed by $ \zeta_u  \zeta_v = \rho Q^2$. 

\par   
In \cite{JC1} a definition of the subtracted TMD quark distribution is given  by taking the light-cone gauge link in the unsubtracted one. The subtracted  one is divided with a different soft-factor than that in Eq.(\ref{SF1}). 
The used soft factor is defined as a combination of different products of four gauge links. 
The light-cone singularity is cancelled by the soft factor. However, 
the soft factor contains gauge links along non-light-cone directions, characterized  by the rapidity $y_n$. 
In the limit $y_n\to \pm \infty$, light-cone singularity appears.  Therefore, the introduced TMD quark distribution in \cite{JC1}, denoted as $q_J(x,b, \zeta_c)$, depends on the parameter $\zeta_c$:
\begin{equation} 
  \zeta_c^2= 2 (x P_A^+)^2 e^{-2 y_n}. 
\end{equation}  
Although the two subtracted distributions $q$ and $q_J$ are different, they have the same soft divergences. 
Therefore, the difference between the two distributions can be calculated perturbatively.  
From explicit results in \cite{SY} we can derive the relation between the two distributions at one-loop:
\begin{equation} 
   q(x, b, \zeta_u, \rho) = C (b, \zeta_u, \rho, \zeta_c) q_J(x, b, \zeta_c) ,  
\label{JJMY}     
\end{equation} 
with the perturbative coefficient as:  
\begin{equation} 
   C (b, \zeta_u, \rho, \zeta_c) = 1 +\frac{\alpha_s C_F} {2\pi}    \biggr [ - \frac{1}{2}  
       \ln^2 \frac{\zeta_u^2}{\mu^2} - \ln \frac{\zeta_u^2}{ \rho \zeta_c^2}    
        \ln\frac{\mu^2 b^2 e^{2\gamma}}{4}   +\ln\frac{\zeta_u^2}{\mu^2} -2  -\frac{\pi^2}{2}  \biggr ] 
          + {\mathcal O} (\alpha_s^2), 
\end{equation} 
where $\gamma$ is Euler constant. 
Such a relation is also expected in the case of TMD fracture functions. 
\par 
The factorization with TMD parton distributions in \cite{JC1} is then given as:
\begin{eqnarray}
W^{\mu\nu} =  - g_\perp^{\mu\nu} \frac{1}{N_c} H_J (Q) \int d^2 b \  e^{-i b\cdot q_\perp} 
   q_J (x, b, \zeta_c ) \bar q_J (y, b, \zeta_{\bar c})  +\cdots .  
\label{TMDDYJC}      
\end{eqnarray} 
In the above the parameter $\zeta_{\bar c}$ in the anti-quark distribution $\bar q_J$,  corresponding 
to $\zeta_c$ in $q_J$,  is not independent. It is given 
as $ 2(y P_B^-)^2 e^{2 y_n}$.  This results in that the perturbative coefficient $H_J(Q)$ does not depend 
on $y_n$ or is free from light-cone singularities. 

\par 
In the factorization given in Eq.(\ref{TMDDY}) the parameter $\zeta_u$ and $\zeta_v$ are independent. 
If we take $\zeta_u =\zeta_c$ and $\zeta_v= \zeta_{\bar c}$,  and express the TMD parton distributions in Eq.(\ref{TMDDY}) with those 
defined in \cite{JC1} by using Eq.(\ref{JJMY}) , we obtain the factorization formula in Eq.(\ref{TMDDYJC}) and  
the perturbative coefficient $H_J(Q)$:
\begin{eqnarray} 
H_J (Q) &=& C(b, \zeta_c, 1, \zeta_c)C(b, \zeta_{\bar c} , 1, \zeta_{\bar c} ) H (Q, \zeta_c, \zeta_{\bar c}) 
\nonumber\\
  &=&  1 + \frac{\alpha_s C_F}{2\pi} \biggr [ 3 \ln\frac{Q^2}{\mu^2} - \ln^2\frac{Q^2}{\mu^2} + \pi^2 -8 \biggr] 
    + {\mathcal O}(\alpha_s^2).   
\end{eqnarray} 
The derived coefficient is in agreement with the coefficient given in \cite{JC1}.  This shows that the two factorizations 
with differently defined TMD parton distributions are equivalent.  It is noted with  $\zeta_u =\zeta_c$ 
and $\zeta_u=\zeta_{\bar c}$ that we have $\rho=1$ and the perturbative coefficient $C$ does not depend on $b$.  

\par 
In the framework of SCET, TMD factorization of Drell-Yan processes has been studied intensively with the so-called $\delta$-regulators in \cite{EIC}.  Although the regulator is introduced in SCET,  in full QCD it is equivalent to take light-cone gauge links to define unsubtracted TMD parton distributions and soft factor. 
But, the used gauge links are modified\cite{ESVI}. E.g.,  the gauge link along the $n$-direction,  or ${\mathcal L}_u$ with $u^+=0$ is modified as: 
\begin{equation} 
  P \exp \biggr ( -i g_s \int^0_{-\infty}  d\lambda n\cdot G(\lambda n +x)\biggr ) \to 
  P \exp \biggr ( -i g_s\int^0_{-\infty} d\lambda n\cdot G(\lambda n +x) e^{ - \delta^+ \lambda} \biggr ).
\end{equation} 
With the small- but finite $\delta^+$ light-cone singularities are regularized.  Because the subtracted 
TMD parton distributions involve a soft factor containing gauge links along the $n$- and $l$-direction, they depend 
on the parameters $\delta^+$ and $\delta^-$, the later regularizes the light-cone divergences in gauge links 
along the $l$-direction. Taking $\delta^+ \propto \delta^-$,  the subtracted TMD parton distributions are free 
from light-cone singularities. This is an advantage of the $\delta$-regulator.  With it TMD parton distribution functions have been calculated at two loop in \cite{ESVpdf}.

\par 
The factorization with TMD parton distributions defined with the $\delta$-regulators takes a similar form 
as discussed in the above.  It is equivalent to the factorization with TMD factorization defined in \cite{JC1}, as shown in detail in \cite{CR1}. Therefore,  TMD factorizations with subtracted TMD parton distributions defined in different ways discussed here are all equivalent.  Although the discussed equivalence is only for TMD parton distributions,  it is expected that there is also such an equivalence in the case of TMD fracture functions. 
In this work we will only study the properties of TMD fracture function defined at the beginning of this section.

\par\vskip10pt
\noindent 
{\bf 2.3. Factorization with integrated fracture function} 
\par 
We consider here the case that the transverse momentum of the lepton pair is large, e.g., $q_\perp\gg \lambda Q$. 
In this case the large $q_\perp$ is generated by hard radiations of partons. Therefore, one can neglect 
transverse momenta of incoming partons. This results in that one can use collinear factorization. 
In this kinematical region, the hadronic tensor at leading power is factorized as:
\begin{eqnarray}
W^{\mu\nu} &=&  \int d x_A d x_B \sum_{a,b} w_{ab}^{\mu\nu} (x_A P_A, x_B P_B, q) 
    F_{a/h_A} (x_A, \xi, k_\perp) f_{b/h_B} (x_B), 
\label{CFFAC}        
\end{eqnarray} 
where the sum is over all possible partons. $w^{\mu\nu}$ can be calculated with perturbation theory. 
$f_{b/B}$ is the standard parton distribution at twist-2, whose definition is given in \cite{PDFDF}. 
$F_{a}$ is the fracture function of $h_A$. $w_{ab}^{\mu\nu}$ can be calculated with perturbation theory.  At leading order they are determined by the tree-level partonic process $a+b\to \gamma^* + c$. 
We will only discuss the quark fracture 
function $F_q$ here, because of that it corresponds to the TMD fracture function introduced in the previous subsection. 

\par 
The quark fracture function appearing in Eq.(\ref{CFFAC})  is defined in \cite{BeSo} as:
\begin{eqnarray} 
  F_{q/h_A}  (x, \xi,k_\perp) =  \int \frac{d \lambda  }{4\pi  } e^{i  x P_A^+ \lambda   } 
     \sum_X \langle h_A (P_A) \vert \bar q (0) {\mathcal L}_n (0) \gamma^+  \vert X, \gamma (k) \rangle 
     \langle \gamma (k),X \vert 
     {\mathcal L}_n^\dagger (\lambda n )  q (\lambda n  ) \vert h_A (P_A) \rangle. 
\label{DEFCF}           
\end{eqnarray}  
It is noted that in the above the transverse momentum of the quark as a parton is integrated over. Because of this, the gauge links along the direction $n^\mu =(0,1,0,0)$  are used. With this definition there can be light-cone singularities 
in different contributions to $F_q$, but the sum is free from the singularities.  
  
\par

\par\vskip20pt
\noindent 
{\bf 3. Factorization of Integrated Fracture Function} 

\par 
In this section we study  the factorization of the integrated fracture function.  
If the observed photon in the final state has large transverse momentum, i.e., $k_\perp \gg  
\Lambda_{QCD}$,  the integrated fracture function has a perturbatively calculable  part.  This part can be 
separated.  
The expected factorization takes the form:  
\begin{equation} 
F_{q/h_A} (x, \xi, k_{\perp}) =\sum_{a} \int \frac{d y }{y }   f_{a/h_A}(y) \biggr  [     
     {\mathcal H}_{a} (x/y ,\xi/y, k_\perp)  + \sum_b \int \frac{dz}{z} D_b (z) {\mathcal H}_{ab} (x/y, \xi/(yz) , k_\perp/\sqrt{z}) \biggr ],   
\label{PAFAC}      
\end{equation} 
where the sum over  $a$ or $b$ is the sum over all possible partons in QCD,   
 including the contributions 
with the parton $a$ as a quark whose flavor is different than that of $q$ in $F_{q/h_A}$. 
${\mathcal H}_a$ and ${\mathcal H}_{ab}$ are perturbative coefficient functions. 
$f_{a/h_A}$ is the twist-2 parton distribution of $h_A$. $D_{a}$ is the fragmentation function 
of a parton $a$ decaying into a photon.  
At one loop level, only ${\mathcal H}_{q,g}$ and ${\mathcal H}_{qg}$ are nonzero. 
The coefficient functions  
${\mathcal H}_{a}$ and ${\mathcal H}_{ab}$,  where the parton $a$ has the flavor other than 
that of $q$,   will become 
nonzero beyond one-loop level. 
 \par  
At tree-level, i.e., at the order ${\mathcal O}(\alpha_s^0)$, only ${\mathcal H}_q$ is nonzero. It is the same 
as the integrated quark fracture function of a quark at the order:  
\begin{eqnarray} 
 F_{q/q}^{(0)}  (x, \xi, k_\perp) =  {\mathcal H}^{(0)}_{q} (x, \xi, k_{\perp}) =  2 e^2 \delta (1-x-\xi)  
 \frac{1}{k_\perp^2} \biggr ( \gamma(\xi)  -\frac{\epsilon}{2} \xi^2 \biggr) .  
 \label{LINF} 
\end{eqnarray} 
The function $ \gamma(\xi)$ is defined as:
\begin{equation} 
    \gamma (\xi) = 1+ (1-\xi)^2, 
\end{equation} 
which will be often used in our work.
At one-loop, the perturbative coefficient functions receive contributions from Fig.2, Fig.3 and Fig.4. 
The one-loop correction of  $ {\mathcal H}_{q}$ is from diagrams given in Fig.2 and Fig.3. 
The diagrams in Fig.4 give contributions to ${\mathcal H}_g$ and ${\mathcal H}_{gq}$. 
In calculations of Fig.2 and Fig.3 we will have U.V. divergences represented by poles in $\epsilon=4-d$. 
Terms with U.V. poles are subtracted. There are collinear- and I.R. divergences. The I.R. divergences will 
be cancelled in the final results.  The collinear divergences will be factorized into $f_{q/h_A}$ so that 
$ {\mathcal H}^{(1)}_{q}$ is finite. Besides these singularities, there are light-cone singularities. They will also be cancelled because the transverse momentum of the parton is integrated here. 

\par
\begin{figure}[hbt]
	\begin{center}
		\includegraphics[width=9cm]{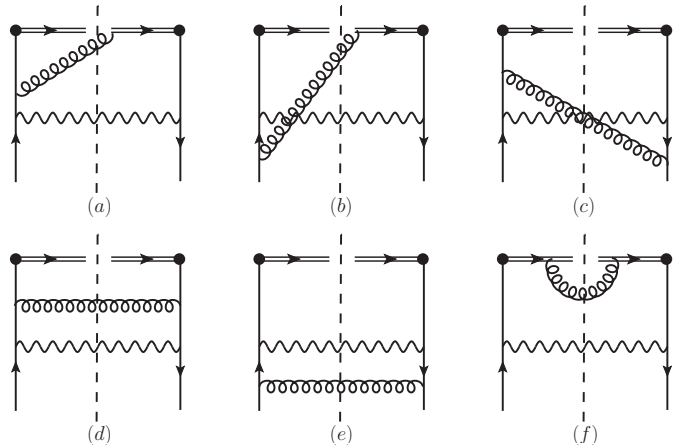}
	\end{center}
	\caption{  Diagrams for the fracture function in the case of large transverse momenta. Double lines represent gauge links. Self-energy diagrams 
	and complex conjugated diagrams should be included.   }
	\label{1LF}
\end{figure}
\par 

\par 
\begin{figure}[hbt]
	\begin{center}
		\includegraphics[width=9cm]{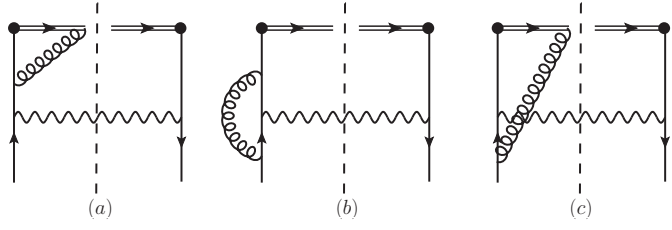}
	\end{center}
	\caption{  One-loop diagrams for the fracture function. Double lines represent gauge links. Self-energy diagrams 
	and complex conjugated diagrams of the diagrams are not given here.     }
	\label{1LV}
\end{figure}

\par 
We first look at the contributions from Fig.2 and Fig.3.  They are one-loop corrections of fracture function 
 $F_{q/q}$ of an initial quark. 
The contributions from those diagrams,  where one gluon is attached to gauge links along the direction $n$, 
have light-cone singularities. But they are cancelled in the sum.  In the sum of the contributions from Fig.2a and Fig.3a and the sum of Fig.2b and Fig.3c there is no light-cone singularity: 
\begin{eqnarray}
F_{q/q} \Big\vert_{2a+3a} &=& \frac{e^2 \alpha_s C_F}{\pi k_\perp^2} \gamma(\xi) \Biggl\{  \left(
\delta(z)+ \frac{x}{\bar \xi(z)_+ }\right) \ln\frac{\mu^2}{k_\perp^2} + \delta(z) (\ln\xi+2)  
+\frac{x}{\bar \xi(z)_+ } \ln\xi\bar\xi^2- \left(\frac{\ln xz}{z}\right)_+ \frac{x }{\bar\xi} \Biggr\}, \nonumber\\
F_{q/q}\Big\vert_{2b+3c} &=& \frac{e^2 \alpha_s C_F}{\pi k_\perp^2} 
\Biggl\{ 
\left(-\frac{2}{\epsilon_c} + \ln\frac{k_\perp^2}{\tilde \mu_c^2} \right) \left[ \delta(z) \left( \gamma(\xi) \ln\bar\xi + 2-\xi \right) + \xi + \frac{\gamma(\xi)}{(z)_+} - 2 \right] \nonumber\\
&&+ \delta(z) \Biggl[ \frac{1}{4} \gamma(\xi) \ln^2\xi + \left(\xi -2 -\frac{3}{2} \gamma(\xi) \ln\bar\xi \right) \ln\xi + \xi^2 \ln\bar\xi + 2 \xi -4 + \frac{\pi^2 \gamma(\xi)}{6} \nonumber\\
&&+ \frac{\gamma(\xi)}{2} {\rm Li}_2(\xi ) + \frac{\gamma(\xi)}{2} {\rm Li}_2\left(-\bar\xi/\xi\right) \Biggr] + \left(\frac{1}{z}\right)_+\left[ \xi^2 +\frac{\gamma(\xi)}{2} \ln\frac{\bar z^2}{x\xi^2\bar\xi^2} \right] + \frac{\gamma(\xi)}{2} \left( \frac{\ln x}{z} \right)_+ \nonumber\\
&&+ \gamma(\xi) \left(\frac{\ln z}{z}\right)_+ + (2-\xi ) \ln\frac{\xi\bar\xi}{z \bar z} \Biggr\}, 
\end{eqnarray}
where $\tilde \mu_c$ is given by $4\pi \mu_c^2 e^{-\gamma}$ with $\mu_c$ as the scale associated with 
the collinear divergence.  $z$ is given by $z=1-x-\xi$. We use the notations $\bar u = 1-u$ to simplify the expressions.
The +-distribution in the above is defined as:
\begin{equation} 
 \int_0^{1-\xi} dx \ t (x)  \frac{1}{(1-x -\xi)_+ }  = \int_0^{1-\xi} dx   \frac{t(x) - t(1-\xi) }{(1-x -\xi) }. 
\label{NPLUS}  
\end{equation}
The contributions from other diagrams have no light-cone divergence. They are: 
\begin{eqnarray}
  F_{q/q}\Big\vert_{3b}  &=&  \frac{e^2 \alpha_s C_F}{\pi k_\perp^2 } \delta(z) 
\left[ \frac{1}{2}\gamma(\xi)\ln\frac{\xi \mu^2}{k_\perp^2} + \xi\bar\xi \left( \frac{2}{\epsilon_c} + \ln\frac{\xi \tilde\mu_c^2}{k_\perp^2} \right) + \frac{3}{2}\xi\bar\xi + 1 \right], 
\nonumber\\ 
 F_{q/q}\Big\vert_{2c}  &= &\frac{e^2  \alpha _s C_F }{\pi  k_{\bot }^2 } 
  \left[        
  \frac{\xi  x}{\bar{z}} \left( \frac{2}{\epsilon _c} + \ln \frac{\tilde{\mu} _c^2}{k_{\bot }^2} \right)
  + \left(\frac{\xi  x}{\bar{z}}+\frac{\bar{x}}{2}\right)   \ln \frac{\xi  x}{z \bar{z}^2}
  +\frac{\bar{x}}{2}  \ln \frac{x z}{\xi  \bar{\xi }^2}
  +\frac{\gamma(\xi)  x+\xi }{\bar{\xi } \bar{z}}
  -\frac{\bar{z}}{\bar{\xi }}   +\xi -x
    \right], 
\nonumber\\     
      F_{q/q} \Big\vert_{2d} &= &\frac{e^2  \alpha _s C_F }{\pi  k_{\perp }^2 \bar{\xi}^2 } 
  \left[        
\gamma(\xi) z \left(  \ln \frac{\mu^2}{k_\perp^2} + \ln \frac{\xi  \bar{\xi }^2}{x z} -1\right)
+\xi  x
  \right], \nonumber\\  
  F_{q/q} \Big\vert_{2e} &= &\frac{e^2  \alpha _s C_F }{\pi  k_{\bot }^2 } 
  \left[        
  -z\left( 1+\frac{x^2}{\bar{z}^2}\right)  \left(    \frac{2}{\epsilon _c} +  \ln \frac{\tilde{\mu} _c^2}{k_{\bot }^2 } +\ln \frac{\xi  x}{z \bar{z}^2} \right)
  +\frac{\xi  \bar{\xi }}{\bar{z}}
  \right].  
 \end{eqnarray}
 The contribution of Fig.2f is zero because of that the gauge links are along the direction $n$ with $n^2=0$. 
Besides the contributions from Fig.2 and Fig.3,  there are corrections from external lines and 
the quark propagator given by:  
\begin{eqnarray} 
F_{q/q} (x,\xi,k_\perp)\biggr\vert_{E} =   -\frac{\alpha_s C_F}{4\pi}  
\biggr [ \biggr ( -\frac{2}{\epsilon_c} + \ln\frac{\mu^2}{\tilde \mu_c^2} \biggr )    
 + 2\biggr ( 1 + \ln\frac{\xi\mu^2}{ k_\perp^2} \biggr )  \biggr ] 
   F_{q}^{(0)}  (x,\xi,k_\perp).  
\end{eqnarray}    
 Summing all contributions, we have the divergent part of the one-loop correction:
 \begin{eqnarray} 
F_{q/q} ^{(1)}\biggr\vert_{div.}   =  \frac{\alpha_s C_F e^2}{\pi k_\perp^2} \biggr (  -\frac{2}{\epsilon_c} \biggr ) 
   \gamma(\xi /y)  \biggr ( \frac{ 1+y^2}{(1-y)_+}  +\frac{3}{2} \delta(1-y)   \biggr ),    
\end{eqnarray} 
with $y=x +\xi$.  The $+$-distribution here is the standard one: 
\begin{equation} 
  \int_{\xi}^1 d y \frac{1}{(1-y)_+} t(y) = \int_{\xi}^1 d y \frac{1}{1-y} ( t(y) - t(1)) + t(1) \ln (1-\xi) . 
\label{STP}  
\end{equation} 
Using the result of the quark distribution of a quark as the target 
\begin{eqnarray} 
 f_{q/q} (x) &=& \delta (1-x) +  \frac{\alpha_s }{2\pi} \left ( -\frac{2}{\epsilon_c} + \ln \frac{ \mu^2}{\tilde \mu_c^2 }\right )  C_F \biggr ( \frac {1+x^2}{(1-x)_+} +\frac{3}{2}\delta (1-x) \biggr ) +{\mathcal O}(\alpha_s^2)
\nonumber\\
      &=&  \delta (1-x) +  \frac{\alpha_s }{2\pi} \left ( -\frac{2}{\epsilon_c} + \ln \frac{ \mu^2}{\tilde \mu_c^2 }
      \right ) P_{qq}(x) +{\mathcal O}(\alpha_s^2), 
\label{qPDF}         
\end{eqnarray}        
we find that the collinear divergent part of the one-loop contribution of $F_q$ from Fig.2 and Fig.3 can be factorized with the quark distribution function. We have then the finite perturbative coefficient:   
\begin{eqnarray}
{\cal H}_q^{(1)} (x,\xi, k_\perp)&=&\frac{e^2  \alpha _s C_F }{\pi  k_{\perp }^2} \left\{
\ln \frac{\mu ^2}{k_{\perp }^2}
 \left[
 \gamma(\xi)\gamma(\bar y/\bar \xi)\left( \frac{1}{(\bar y)_+} -\delta(\bar y)\ln\bar \xi\right) 
 - \gamma(\xi/y)\frac{1+y^2}{(\bar y)_+}
   \right]  \right . 
  \nonumber \\
&&
+\delta (\bar{y}) 
\left[
\gamma(\xi) \left(
-3 \ln\xi \ln\bar{\xi }+2 \ln \bar{\xi }+\frac{ 1}{2} \ln^2\xi
+{\rm Li}_2\left(-\frac{\bar{\xi }}{\xi }\right) + {\rm Li}_2 ( \xi) +\frac{\pi ^2}{6}-4
  \right)
  \right.
\nonumber \\ &&
  -4 \bar{\xi} \left(\ln \bar{\xi }-2  \right)+\xi^2+\xi
\bigg]
+\gamma(\xi) \left(\frac{1}{\bar{y}}\right)_+ 
\left(
2\frac{ \bar{y}}{\bar{\xi }} \ln \frac{x }{\xi  \bar{\xi }^2}
-\ln \frac{x^2}{y^2 \bar{\xi }^2}
+\frac{2 \xi^{2}\left(x y^{2}-\bar{\xi}\right)}{y^{2} \bar \xi \gamma(\xi)}
\right)
\nonumber \\ &&
+ \left(-\gamma(\xi)\frac{ \bar{y}}{\bar{\xi }^2}+\bar{y}+\xi \right) \ln \frac{\bar{y} x}{\xi 
   \bar{\xi }^2}
+\left(2 y-\xi -(y+1)\frac{ x^2}{y^2}\right) \ln \frac{\xi  x}{y^2 \bar{y}}
 -2 \bar{\xi } \ln \frac{y}{\xi  \bar{\xi }}
   \nonumber \\ &&
 \left .
+\frac{2 }{\bar{\xi }}\ln \bar{y}-2 \ln \frac{y \bar{y}}{\xi  \bar{\xi }}
   +\frac{\xi^2-\xi  \bar{\xi }}{y}+\frac{5 \bar{y}+3}{\bar{\xi }}
  +\frac{( \xi ^2-3) \bar{y}}{\bar{\xi }^2}+\frac{\xi
   ^2}{y^2}-3
 \right\} + {\cal O}(\alpha_s^2),
 \end{eqnarray}
 which is finite. 
\par
\begin{figure}[hbt]
	\begin{center}
		\includegraphics[width=13cm]{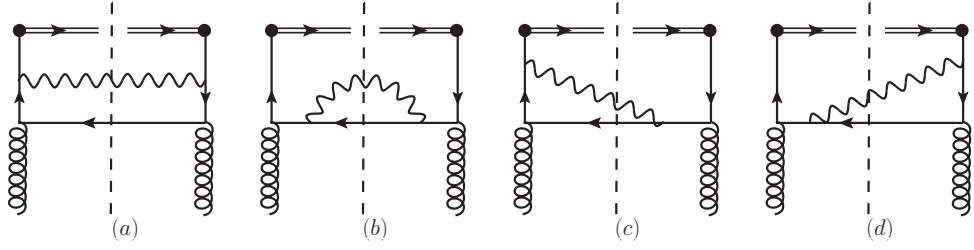}
	\end{center}
	\caption{  One-loop diagrams for the fracture function of a gluon.   }
	\label{1LFG}
\end{figure}
\par
\par 
The contributions from Fig.4  are those to the fracture function $F_{q/g}$ of an initial gluon.  They are U.V. finite.  They only contain collinear divergences represented 
by poles in $\epsilon_c =4-d$. The results are:  
\begin{eqnarray}
 F_{q/g} \bigg\vert _{4a}&=&\frac{e^2  \alpha _s  }{ 2\pi  k_{\perp }^2}
\left[
-\left(\bar{z}^2+z^2\right) \left(\frac{x^2}{\bar{z}^2}+1\right)
\left(
\frac{2}{ \epsilon_c} + \ln\frac{\tilde \mu_c^2}{k_\perp^2}+\ln \frac{\xi  x}{z \bar{z}^2}
\right)
+\frac{\xi  \bar{x}}{\bar{z}}+4 x z
\right] ,
\nonumber \\
F_{q/g} \bigg\vert _{4b}&=&\frac{e^2  \alpha _s  }{ 2\pi  k_{\perp }^2}
\left[
-\left(\bar{x}^2+x^2\right) \left(\frac{z^2}{\bar{x}^2}+1\right)
\left(
\frac{2}{ \epsilon_c} + \ln\frac{\tilde\mu_c^2}{k_\perp^2}+\ln \frac{\xi ^3 x}{z \bar{x}^2}
\right)
+\frac{\xi  }{\bar{x}}\left(\xi +x^2\right)+4 x \bar{x}-3 \xi  x
\right] ,
\nonumber \\ 
 F _{q/g} \bigg\vert _{4c}&=&\frac{e^2  \alpha _s  }{ 2\pi  k_{\perp }^2}
\left[
\left(x^2+z^2\right) \ln \frac{x z}{\bar{x} \bar{z}}-\frac{\left(-\bar{\xi }+x^2+z^2\right)
   \left(\bar{\xi }-2 x z\right)}{\bar{x} \bar{z}}
\right].
\label{IFIG3} 
\end{eqnarray} 
The contribution from Fig.4d is the same as that of Fig.4c. 
There are collinear divergences in the contribution from Fig.4a and 4b. 
 It is clear that the collinear divergence 
in Fig.4b is factorized into the antiquark fragmentation function into the photon, which is the same as the 
quark fragmentation function.  This fragmentation function can be easily calculated. Or  it can be extracted  from the quark fragmentation function into a gluon which 
is well-known. We have:  
\begin{equation} 
  D_q (z)  = \frac{ e^2  }{8 \pi^2  } \left ( -\frac{2}{\epsilon_c} + \ln \frac{\mu^2}{\tilde \mu_c^2 }\right )   \frac {1+ (1-z)^2}{z} + {\mathcal O}(\alpha_s) .   
\end{equation} 
Therefore, we determine $ {\mathcal H}_{gq}$ at the leading order as:
\begin{eqnarray} 
 {{\cal H}}_{gq} (x,\xi, k_\perp)&=&\frac{4\pi\alpha _s}{k_{\perp }^2}\left(x^2+\bar{x}^2\right)\bar{x}  \delta\left(\bar{x} -\xi \right) + {\cal O}(\alpha_s^2). 
\end{eqnarray}    
Using the result of the quark distribution function of a gluon, which is 
\begin{equation} 
  f_{q/g} (x) =\frac{\alpha_s }{2\pi} \left ( -\frac{2}{\epsilon_c} + \ln \frac{ \mu^2}{\tilde \mu_c^2 }\right ) P_{qg} (x) + {\mathcal O}(\alpha_s^2),  \quad   P_{qg} (x) = \frac{1}{2} \biggr ( x^2 + (1-x)^2  \biggr ) 
\end{equation}
at the leading order, we find that the collinear divergence in Fig.4a is factorized.  Hence, we have the finite result 
for ${\mathcal H}_g$ at the leading order:
\begin{eqnarray}
 {\cal H}_g (x,\xi,  k_\perp) &=&\frac{e^2  \alpha _s  }{2 \pi  k_{\bot }^2} \left\{-(y^2+\bar{y}^2 )\left[ 
   \left(\ln \frac{\mu ^2}{k_{\bot
   }^2}+\ln \frac{\xi  x}{\bar{y}
   y^2}\right)
   \left(\frac{x^2}{y^2}+1\right)+\frac{\xi ^2}{y^2}  \right]
   \nonumber \right . \\ 
 &&  -\left(\bar{x}^2+x^2\right) \left(\frac{\bar{y}^2}{\bar{x}^2}+1\right)
   \left(
\ln\frac{\mu ^2}{k_{\bot }^2}
+\ln \frac{\xi ^3 x}{\bar{y} \bar{x}^2}
\right) + 2\left(x^2+\bar{y}^2\right) \ln \frac{x
   \bar{y}}{\bar{x} y}  
   \nonumber  \\ &&\left . 
   -2\left(x^2+\bar{y}^2-\bar{\xi}\right) \frac{\bar{\xi }-2 x
   \bar{y}}{\bar{x} y}
+ \xi \left(  \frac{ y}{\bar{x}}+\frac{ \bar{x}}{y} \right)+8 x \bar{y}
\right\} + {\mathcal O}(\alpha^2_s). 
\end{eqnarray}
\par 

\par  
From our result we can also derive the $\mu$-evolution of the integrated fracture function. It is given by  
\begin{equation} 
 \frac {\partial F_{q/h_A}  (x, \xi, k_\perp, \mu)}{ \partial \ln \mu } = \frac{\alpha_s}{\pi}  \int_{x}^{1-\xi}  \frac{ d y}{y}  P_{qq} ( x/y) F_{q/h_A}  (y, \xi, k_\perp, \mu)  + \cdots. 
\label{EVOIF}     
\end{equation}     
The explicitly given part is the same as the evolution of the standard quark distribution. Only the integration range is different because of that $F_q$ is zero for $x > 1-\xi$.  In Eq.(\ref{EVOIF}) $\cdots$ denotes the contribution from mixing 
of gluon fracture function, which we can not derive from our existing results.  
\par 
Similarly, the factorization in Eq.(\ref{PAFAC}) also holds if we replace the photon with a hadron. 
In this case,  only contributions with fragmentation functions exist because of that the hadron can only be produced through parton fragmentation.  

\par\vskip20pt 
\noindent 
{\bf 4. Factorization of TMD Fracture Function} 
\par\vskip5pt 
\par\vskip10pt 
\noindent 
{\bf 4.1 Factorization at Large Transverse Momenta} 
\par\vskip5pt  

In general TMD fracture function is nonperturbative. However, if 
the transverse momentum $k_{A\perp}$ of the parton  is much larger than $\Lambda_{QCD}$,  
then the $k_{A\perp}$-behavior of the TMD fracture function can be predicted by perturbative QCD.  
The predictions take different factorizations in the cases where the transverse momentum $k_\perp$ 
of the photon is at different orders.  We will study the cases with $k_\perp \gg \Lambda_{QCD}$ and 
$k_\perp \sim \Lambda_{QCD}$.

\par
We consider the case $k_{A\perp},  k_\perp \gg \Lambda_{QCD}$ and $\ell_\perp =k_{A\perp}+ k_\perp \gg \Lambda_{QCD}$. In this case 
there must be at least one  energetic parton in the intermediate state. At leading order of $\alpha_s$, 
the dependence is determined by the process that a primary parton from $h_A$ emits one parton and one photon 
at tree-level before participating in hard scattering. The primary parton has only the transverse momentum at order of $\Lambda_{QCD}$ which can be neglected.  Therefore, in the case $k_{A\perp},  k_\perp \gg \Lambda_{QCD}$ and $\ell_\perp =k_{A\perp}+ k_\perp \gg \Lambda_{QCD}$ one expects
a factorization of ${\mathcal F}_{q/h_A}$ which takes a similar form as that in Eq.(\ref{PAFAC}) with perturbative coefficient functions depending on transverse momenta.   
At the leading order of $\alpha_s$ the factorization is: 
\begin{eqnarray} 
{\mathcal F}_{q/h_A}  (x, k_{A\perp},  \xi, k_{\perp})   &=&   \int_{x+\xi}^1 \frac{d y }{y }  \biggr [ 
 f_{q/h_A}  (y)  {\mathcal C}_{q}  (x/y, k_{A\perp},  \xi/y,   k_\perp)  + f_{g/h_A}  (y){\mathcal C}_g (x/y, k_{A\perp}, \xi/y,  k_\perp)    \biggr ],  
\end{eqnarray}          
where ${\mathcal  C}_{q,g}$ are perturbative coefficient functions. 
Beyond the leading order,  there are other contributions.

\par 
At leading order of $\alpha_s$, ${\mathcal  C}_{q,g}$ receives contributions from Fig.2 and Fig.4 respectively.  The contributions from Fig.2 are essentially those of the fracture function of an initial quark.   
The calculations are straightforward.  The results of each diagram in Fig.2 in momentum space are:
\begin{eqnarray} 
{\cal F}_{q/q} \big\vert_{2a}  &=& \frac{\alpha_s C_F e^2}{2\pi^2 k_\perp^2 D_d} \biggl[ \frac{k_\perp^2 z (1 -x/\xi)}{  D_{\zeta }} -\frac{z \hat \zeta_u^2}{D_{\zeta }} (z\bar\xi + 2z/\bar\xi - 2\bar\xi^2 - 2) + \bar\xi(D_d/D_{\zeta }-1)\biggr] + {\cal O}(\epsilon), 
\nonumber\\
{\cal F}_{q/q} \big\vert_{2b}  &=& \frac{\alpha_s C_F e^2}{2\pi^2 k_\perp^2 \ell_\perp^2} \biggl[ \frac{z \hat \zeta_u^2}{D_{\zeta }} \left( \frac{k_\perp^2 z \left(\xi ^2+x^2+\xi  x+x\right)}{\xi  \bar\xi  D_d} -\xi  x+x+1 \right) -\frac{k_\perp^2 x z}{\xi  D_d} 
\nonumber\\
&& \qquad\qquad+ \frac{\ell_\perp^2}{\xi D_d D_\zeta} \left(\xi  D_d+k_\perp^2 x z + z\zeta_u^2 \xi  (z- x\bar\xi-1)\right) - 1  \biggr] + {\cal O}(\epsilon), 
\nonumber\\
{\cal F}_{q/q} \big\vert_{2c} &=& \frac{\alpha_s C_F e^2}{2\pi^2 k_\perp^2 D_d} \biggl[ \frac{\xi ^2 D_c (x-\xi  \bar x)+k_\perp^2 z \left(\xi -3 x^2+2 x-3 x\xi\right)}{\xi  \bar\xi^2 D_d}-\frac{k_\perp^4 x z^2 (\xi +x)}{\ell_\perp^2 \xi ^2 \bar\xi^2 D_d} -\frac{k_\perp^2 z \left(\xi -x^2+\xi  x\right)}{\ell_\perp^2 \xi  \bar\xi} 
\nonumber\\
&&\qquad\qquad + \frac{\ell_\perp^2 (\xi -2 \xi  x)}{\bar\xi^2 D_d}+\frac{\bar x D_d}{\ell_\perp^2}-\frac{\bar x}{\bar\xi}  \biggr] + {\cal O}(\epsilon), 
\nonumber\\
{\cal F}_{q/q} \big\vert_{2d} &=& \frac{\alpha_s C_F e^2}{\pi^2 \bar\xi^2 D_d^2} \biggl[ \frac{\ell_\perp^2 ((\xi -2) (\xi +x)+2)}{k_\perp^2} -\frac{ D_c \xi\left(\xi ^2-2 \xi -\bar\xi x+2\right)}{k_\perp^2} + \frac{z \left(\xi +z^2+z\right)}{\xi } \biggr] + {\cal O}(\epsilon), 
\nonumber\\
{\cal F}_{q/q} \big\vert_{2e}  &=& \frac{\alpha_s C_F e^2 z}{\pi^2 \bar\xi \ell_\perp^2 D_d} \biggl[ \frac{k_\perp^2 x^2 z}{\xi ^2 \bar\xi D_d}+\frac{\ell_\perp^2 x (x-z)}{\xi  \bar\xi D_d}+\frac{2 x^2 -\xi  \bar\xi +2 \xi  x-x}{\xi } \biggr] + {\cal O}(\epsilon), 
\nonumber\\
{\cal F}_{q/q} \big\vert_{2f}  &=& -\frac{2\alpha_s C_F e^2}{\pi^2k_\perp^2} \cdot \frac{z\hat \zeta_u^2 \gamma  (\xi)}{ (\ell_\perp^2 + z^2\hat \zeta_u^2) ^2} + {\cal O}(\epsilon).
 \label{RF2A} 
  \end{eqnarray} 
In the above we have used the notations: 
\begin{equation} 
 \hat \zeta_u^2 = 2\frac{u^-}{u^+} (P_A^+)^2, \quad  \ell_\perp = k_{A\perp} + k_\perp, \quad D_c =\biggr ( \ell_\perp + \frac{z k_\perp}{\xi} \biggr )^2, 
     \quad  D_d = \biggr ( \ell_\perp - \frac{z k_\perp}{1-\xi} \biggr )^2 + \frac{ x z k_\perp^2 }{\xi (1-\xi)^2}. 
\end{equation} 
The contributions involving gauge links are from Fig.2a, 2b and 2f. 
 For them we need to take the limit $\hat \zeta_u \to \infty$ according to our definition in Eq.(\ref{DEFFF}). 
From Eq.(\ref{RF2A}) the contributions contain these terms  
\begin{equation} 
\frac{ \hat \zeta_u^2 z^2 }{(\ell_\perp^2 + z^2 \hat \zeta_u^2)}, \quad  \frac{ \hat \zeta_u^2 z }{(\ell_\perp^2 + z^2\hat \zeta_u^2)},  
\quad \frac{ \hat \zeta_u^2 z }{(\ell_\perp^2 + z^2 \hat \zeta_u^2)^2},  
\end{equation} 
which contain the parameter $\hat \zeta_u$. We should take these terms as distributions and take the limit. The results of the 
limit are: 
\begin{eqnarray} 
\frac{ \hat \zeta_u^2 z^2 }{(\ell_\perp^2 + z^2 \hat \zeta_u^2)} &=& 1  + {\mathcal O}( \hat \zeta_u^{-2}), \quad 
\frac{ \hat \zeta_u^2 z }{(\ell_\perp^2 + z^2 \hat \zeta_u^2)^2} = \frac{1}{2 \ell_\perp^2} \delta (z) + {\mathcal O}(\hat \zeta_u^{-2}), 
\nonumber\\
\frac{ \hat \zeta_u^2 z }{(\ell_\perp^2 + z^2 \hat \zeta_u^2)} &=& \frac{1}{(1-x -\xi)_+ } 
  +\frac{1}{2} \delta (z) \ln\frac{ x^2 \hat \zeta_u^2}{\ell_\perp^2}   + {\mathcal O}(\hat \zeta_u^{-2}), 
\end{eqnarray}
with the $+$-distribution defined as given in Eq.(\ref{NPLUS}). 
In taking the limit, we should make the following arrangement for the term with $\hat\zeta_u^2$ combined with $D_d$:
\begin{equation} 
\frac{ \hat \zeta_u^2 z }{(\ell_\perp^2 + z^2 \hat \zeta_u^2) D_d}  =\frac{ \hat \zeta_u^2 z }{(\ell_\perp^2 + z^2 \hat \zeta_u^2)}
\frac{1}{\ell_\perp^2} +\frac{\hat  \zeta_u^2 z }{(\ell_\perp^2 + z^2 \hat \zeta_u^2)} \biggr (  \frac{1}{D_d} -\frac{1}{\ell_\perp^2} \biggr ), 
\end{equation} 
where the last term is finite at $z=0$ in the limit $\hat \zeta_u \to \infty$. This is important for the cases when 
we integrate over $\ell_\perp$ later. Otherwise, we will have spurious divergence at $z=0$. 
After taking the limit, we  obtain ${\mathcal C}_q$ from Fig.2: 
\begin{eqnarray}
{\mathcal C}_q (x, k_{A\perp}, \xi,  k_\perp ) 
&=& \frac{\alpha_s C_F e^2}{\pi^2 k_\perp^2} 
\Biggl[ \frac{\delta(z) \gamma(\xi) }{\ell_\perp^2} \left(\ln \frac{\zeta_u^2}{\ell_\perp^2} -1\right) + \frac{\gamma(\xi) /z_+ -x-\bar\xi}{\ell_\perp^2} - \frac{x(\xi^2\ell_\perp^2 + z^2 k_\perp^4/\ell_\perp^2)}{\xi\bar\xi^2 D_d^2} \nonumber\\
&&\hspace{1cm} + \frac{\gamma(\xi) \bar\xi/z_+ - x-1}{\bar\xi D_d} + \frac{k_\perp^2(\gamma (\xi) -z^3+3 z^2+\xi  z-4 z)}{\xi\bar\xi \ell_\perp^2 D_d} \Biggr] + {\mathcal O}(\epsilon), 
\label{Cq}  
\end{eqnarray} 
where we have neglected terms at higher orders of $\epsilon=4-d$.        
It is interesting to note that there is no light-cone singularity in the contribution from Fig.2b, i.e., it does not contain terms with $\ln \zeta_u^2$.  Only the contribution from Fig.2a has terms with  $\ln \zeta_u^2$.  
This fact leads to that the Collins-Soper equation of the TMD fracture function, which governs the $\zeta_u$-dependence, is the same as that of TMD quark distribution. This is within expectation.

The contributions to ${\mathcal C}_g$ are from Fig.4.  The diagrams in Fig.4 are essentially the TMD quark fracture function of a gluon as target.  We obtain: 
\begin{eqnarray} 
{\cal F}_{q/g} \big\vert_{4a} &=& \frac{e^2 \alpha_s z}{2\pi^2 \xi \bar\xi D_d \ell_\perp^2} 
\Bigg [   \frac{x}{\xi \bar\xi D_d }  \left(x z^{2} k_{\perp}^{2}-\xi \left(\overline{z}^{2}+x^{2}-x z+z^{2}\right)\ell_{\perp}^{2} \right) 
+2(\overline{z}+1)\left(x^{2}+z^{2}\right)
\nonumber\\
    &&  -3\left(x^{2}+z\right)+1\Bigg]+ {\cal O}(\epsilon), 
\nonumber\\
{\cal F}_{q/g} \big\vert_{4b} &=& \frac{e^2 \alpha_s z}{2\pi^2 \xi^2 \bar\xi D_d D_c} 
\Bigg[  \frac{-x}{\xi \bar\xi D_d }  \left(x z^{2} k_{\perp}^{2}+\xi^{2} D_{c}\left(\bar{\xi}^{2}+\bar{x}-x(\xi+4 z)\right)\right) 
+\bar{x}\left(-x \bar{\xi}+4 x^{2}+z^{2}\right)
\nonumber\\
  && +x^{3}-(\xi+2) 
x+1\Bigg]+ {\cal O}(\epsilon), 
\nonumber\\
{\cal F}_{q/g} \big\vert_{4c} &=& {\cal F}_q\big\vert_{3d} =\frac{e^2 \alpha_s z}{4\pi^2\xi^{2} \bar{\xi} D_{c} D_{d} \ell_{\perp}^{2}}
\Bigg[  \frac{x z \bar{x} \bar{z} k_{\perp}^{2}\left(\ell_{\perp}^{2}-\xi D_{c}\right)}{\xi \bar{\xi} D_{d}}
+x \xi\left(3 \bar{\xi}^{2}-8 x z\right)\frac{D_{c} \ell_{ \perp}^{2}}{\bar{\xi} D_{d}}
+z \bar{\xi} D_{d}\left(x^{2}+z^{2}\right)
\nonumber\\
  && -\xi D_{c}\left(\xi \bar{\xi}^{2}+x\left(x(x-\bar{x})-\xi^{2}-4 \xi z\right)\right)
+\ell_{\perp}^{2}\left(x^{2}(\bar{z}+x)-(\bar{\xi}-2 x)^{2}-x\right)\Bigg]+ {\cal O}(\epsilon).  
 \label{RF3A} 
  \end{eqnarray}  
These results do not depend on $\hat \zeta_u$. From them we have:  
\begin{eqnarray}
{\mathcal C}_g (x,  k_{A\perp}, \xi, k_\perp )  &=&  \frac{\alpha_s e^2 z}{2\pi^2} 
\Biggl[ \frac{x z k_{\perp }^2-2 \xi ^2 x D_c}{\xi^2 \bar\xi^2 D_c D_d^2} + \frac{2x^2(1+z)-\xi ^2 x-3 x+1}{\xi^2 \bar\xi D_c D_d} + \frac{z (\bar\xi^2-2xz)}{\xi ^2 D_c \ell_\perp^2} \nonumber\\
&&\hspace{1cm} -\frac{2x^2(z-\xi) -\xi^3 -3 \xi ^2 x-x}{\xi\bar\xi  D_d \ell_\perp^2}-\frac{x z k_{\perp }^2}{\xi \bar\xi^2 D_d^2 \ell_\perp^2} \Biggr] + {\mathcal O}(\epsilon), 
\end{eqnarray}
 where we have only kept the contributions at the leading order of $\epsilon$.
 
 \par 
In the case that $k_\perp, k_{A\perp}$ and $\ell_\perp$ are all much larger than $\Lambda_{QCD}$,  the $k_\perp$- and $k_{A\perp}$- dependence are completely determined by perturbative coefficients ${\mathcal C}_q$ and ${\mathcal C}_g$.  It is possible that one has $k_\perp, k_{A\perp}\gg \ell_\perp\sim \Lambda_{QCD} $.  In this case the TMD fracture function can be factorized with TMD quark distribution. 
By using the leading order result it is easy to find: 
\begin{equation} 
 { \mathcal F}_{q /h_A} (x,  k_{A\perp},  \xi, k_\perp) = \int d k^2_{q\perp} \frac{ dy}{y} C_{\perp} ( x/y, k_{A\perp}, \xi,  k_{\perp}, k_{q\perp})  f_{q/h_A}  (y, k_{q\perp} ), 
\end{equation} 
with $C_{\perp}$ determined as:
\begin{equation} 
   C_\perp ( x/y, k_{A\perp}, \xi,  k_{\perp}, k_{q\perp}) =  2 e^2 \delta (1-x-\xi) \delta^2 (k_{A\perp} + k_\perp - k_{q\perp} )   
 \frac{1}{k_\perp^2}  \gamma (\xi)  + {\mathcal O}(\alpha_s). 
\end{equation}  
Beyond the leading order here the parton fragmentation functions of a photon will be involved.

\par 
With the given results of Fig.2 and Fig.4 we can take the limit 
$k_{A\perp} \gg k_\perp \sim \Lambda_{QCD}$ 
to find the factorization in this limit.  It is straightforward to find that the contributions from Fig.4 
are proportional to $1/k_{A\perp}^{4}$, while the contributions from Fig.2 are proportional to 
$1/k_{A\perp}^{2}$.  Therefore, we can neglect the contribution from Fig.4.  The contribution
from Fig.2 in the limit reads: 
\begin{equation} 
{\mathcal F}_{q/q} (x, k_{A\perp}, \xi, k_\perp)\biggr\vert_{Fig.2 }
= \frac{e^2 \alpha_s C_F}{ \pi^2 k_\perp^2 k_{A\perp}^2} \frac{\gamma(\xi ) }{\bar \xi} 
\biggr [ \delta (1-x/\bar\xi ) \biggr ( \ln \frac{\zeta_u^2}{k_{A\perp} ^2}   -1 \biggr ) 
  + \frac{2}{ (1-x/\bar \xi)_+} -1 - x/\bar \xi \biggr ].  
\end{equation}   
Comparing the leading order result of the integrated fracture function in Eq.(\ref{LINF}), we find the 
factorization in the limit $k_{A\perp} \gg k_\perp \sim \Lambda_{QCD}$ at the leading order of $\alpha_s$ as:
 \begin{equation} 
{\mathcal F}_{q/h_A}  (x, k_{A\perp}, \xi, k_\perp) = \int_x^{1-\xi} \frac{dy}{y} C_q(x/y, k_{A\perp}, \zeta_u) 
   F_{q/h_A}  (y, \xi, k_\perp), 
\end{equation} 
with the perturbative coefficient function given by:
\begin{equation} 
C_q (x, k_{A\perp}, \zeta_u) = \frac{\alpha_s C_F} {2\pi^2 k_{A\perp}^2} \biggr [ \delta (1-x ) \biggr ( \ln \frac{\zeta_u^2 }{k_{A\perp}^2}   -1 \biggr ) 
  + \frac{2}{ (1-x)_+} -1 - x \biggr ]  +{\mathcal O} (\alpha_s^2).        
\end{equation} 
In this case only the $k_{A\perp}$-dependence is determined by perturbative QCD, as expected. The $+$-distribution here is the standard one.

\par\vskip20pt
\noindent 
{\bf 4.2. Factorization in Impact-Parameter Space} 
\par 
In this section we study factorization of the fracture function in $b$-space. 
We consider the limit $b\to 0$. In this limit, the function can be factorized into different forms, depending 
on the order of $k_\perp$. We first look at the case that $k_\perp \gg \Lambda_{QCD}$.  In this case, 
the dependence on $k_\perp$ can be calculated with perturbative theory. The expected form of the 
factorization reads: 
\begin{eqnarray}
  {\mathcal F}_{q/h_A}  (x, b, \xi, k_{\perp})  &=&   \int_{x+\xi}^1  \frac{d y }{y }   \sum_a f_{a/h_A}  (y)    
     \hat {\mathcal H}_{a} (x/y ,  b,  \xi/y,   k_\perp)
\nonumber\\      
 &&  +\sum_{ab} \int \frac{dy}{y} \int \frac{ dz}{z}  f_{a/h_A}  (y) D_b (z)     
     \hat {\mathcal H}_{ab } (x/y , b, \xi/(yz) ,    k_\perp/\sqrt{z} )   + {\mathcal O} (b).  
\label{FACFFB}     
\end{eqnarray}
The factorization form is similar to that given in Eq.(\ref{PAFAC}).  At one-loop level, only the perturbative 
coefficient functions $ \hat {\mathcal H}_{q,g}$ and $ \hat {\mathcal H}_{qg}$ are nonzero. Others become 
nonzero beyond one-loop. 
To determine perturbative coefficient functions, we note that these  functions do not depend on the initial hadrons. They are determined by fracture functions of an initial parton instead of a hadron, i.e.,  fracture distributions of a parton. 
Therefore, we need to calculate ${\mathcal F}_{q/q}$ and   ${\mathcal F}_{q/g}$.  At tree-level, we have ${\mathcal F}_{q/q}$ 
and hence $ \hat {\mathcal H}_{q}$ as: 
\begin{eqnarray}  
    {\mathcal F}_{q/q}  (x, b, \xi, k_{\perp}) = \hat {\mathcal H}_{q}  (x, b, \xi,  k_{\perp}) = 2 e^2 \delta (1-x-\xi)  
 \frac{1}{k_\perp^2} \biggr ( \gamma (\xi)  -\frac{\epsilon}{2} \xi^2 \biggr) + {\mathcal O}(\alpha_s)
     + {\mathcal O}(b). 
 \label{TREEFB}  
\end{eqnarray}
Other perturbative coefficient functions are zero at tree-level. They become nonzero at higher orders of  $\alpha_s$.

The one-loop correction to the tree-level ${\mathcal F}_q$ is represented by diagrams in Fig.2 and Fig.3. The correction from Fig.2 is the real part, while the correction from Fig.3 is the virtual part. The virtual part in the transverse momentum space is proportional to $\delta^2 (\ell_\perp)$ as the tree-level result does. 
We obtain the virtual part of each diagram after the subtraction of U.V. poles in the $b$-space as: 
\begin{eqnarray}
{\mathcal F}_{q/q}\Big\vert_{3a} &=& -\frac{e^2 \alpha_s C_F}{2\pi} \delta(z) \gamma(\xi) \frac{1}{k_\perp^2} 
\left[ \ln^2\frac{k_\perp^2}{\xi \zeta_u^2} + \ln\frac{k_\perp^2}{\xi \mu^2} + \ln\frac{k_\perp^2}{\xi \zeta_u^2} + \frac{4\pi^2}{3} - 2 \right], \nonumber\\
{\mathcal F}_{q/q}\Big\vert_{3b} &=& \frac{e^2 \alpha_s C_F}{\pi} \delta(z) \frac{1}{k_\perp^2} 
\left[ \frac{1}{2}\gamma(\xi)\ln\frac{\xi \mu^2}{k_\perp^2} + \xi\bar\xi \left( \frac{2}{\epsilon_c} + \ln\frac{\xi \tilde\mu_c^2}{k_\perp^2} \right) + \frac{3}{2}\xi\bar\xi + 1 \right], \nonumber\\
{\mathcal F}_{q/q} \Big\vert_{3c} &=& \frac{e^2 \alpha_s C_F}{\pi} \delta(z) \frac{1}{k_\perp^2}
\Biggl\{ -\frac{2\gamma(\xi)}{\epsilon_c^2} + \frac{1}{\epsilon_c} \left[ \gamma(\xi)\ln\frac{\zeta_u^2}{x^2 \tilde \mu_c^2} + \xi^2+2\xi-4 \right] \nonumber\\
&&+ \frac{1}{4}\gamma(\xi) \left(\ln^2\frac{\zeta_u^2}{ x^2 k_\perp^2} -\frac{1}{2}\ln^2\frac{\zeta_u^2}{x^2 \tilde\mu_c^2} + \ln^2\frac{k_\perp^2}{\tilde\mu_c^2} \right) 
+ \frac{1}{8}\gamma(\xi) \ln\frac{\zeta_u^2}{x^2\tilde\mu_c^2} \left(\ln\frac{\zeta_u^2}{x^2 k_\perp^2} - 3 \ln\frac{k_\perp^2}{\tilde\mu_c^2} \right) \nonumber\\
&& -\frac{1}{4} \left(\xi^2 - 2\gamma(\xi) \ln\xi\bar\xi^2 - 3\xi +2 \right) \ln\frac{\zeta_u^2}{x^2 k_\perp^2} -\frac{1}{4} \left(\xi ^2 - 2\gamma(\xi) \ln\xi + 3\xi -2\right) \ln\frac{\zeta_u^2}{ x^2 \tilde\mu_c^2} \nonumber\\
&& -\frac{1}{4} \left(\xi ^2 + 2\gamma(\xi) \ln\xi + \xi -6\right) \ln\frac{k_\perp^2}{\tilde\mu_c^2} + \frac{\gamma(\xi)}{2} \ln^2\xi\bar\xi - \gamma(\xi) {\rm Li}_2\left(-\xi/\bar\xi\right) \nonumber\\
&& -\frac{\pi^2 \gamma(\xi)}{24} + (\xi-2)(2+\ln\xi) \Biggr\}, \nonumber\\
{\mathcal F}_{q/q} \Big\vert_{E} &=& -\frac{e^2 \alpha_s C_F}{2\pi} \delta(z) \frac{1}{k_\perp^2}
\Biggl\{ \gamma(\xi) \Biggl[ \left( -\frac{2}{\epsilon_c} + \ln\frac{\mu^2}{\tilde \mu_c^2} \right) - 2\left( -\frac{2}{\epsilon_c} + \ln\frac{\mu^2}{\tilde \mu_c^2} \right) \nonumber\\
&&+ 2\left( 1 + \ln\frac{\xi \mu^2}{k_\perp^2} \right) \Biggr] - \xi^2 \Biggr\}, 
\label{1LV}     
\end{eqnarray}
where the poles given as $1/\epsilon_c$ with $\epsilon_c =4-d$ are either collinear ones or I.R. ones. $\mu$ is the renormalization scale from the U.V. subtraction. There is a term with the double pole from Fig.3c.  This double pole consists of 
a collinear pole and an I.R. pole.  $\tilde b^2$ is given by $ b^2 e^{2\gamma}/4$. 
In Eq.(\ref{1LV}) the last two lines stand for the sum of the virtual corrections from corrections of quark propagator, external quark lines and 
the self-energy of gauge links.    
\par 
The real part of one-loop corrections is given by Fig.2. The result of Fig.2 in the transverse momentum space 
is essentially given in Eq.(\ref{RF2A}). The transformation into $b$-space is tedious. We give the result of each diagram of Fig.2 for small $b$: 
 \begin{eqnarray}
{\mathcal F}_{q/q}\Big\vert_{2a} &=&\frac{e^2  \alpha _s C_F }{\pi  k_{\perp }^2} \left\{ 
-\frac{2 \gamma(\xi) }{\epsilon _c^2} \delta (z)
+\frac{\gamma(\xi)}{\epsilon _c} \delta (z) \left(
\ln\frac{\tilde{\mu} _c^2}{\zeta _u^2 }   -2 \ln \frac{\tilde{\mu} _c^2\xi}{k_{\perp }^2} + 1-2\frac{ \bar{\xi }}{\gamma(\xi)}  \right) \right .
\nonumber \\ &&
-\frac{\delta(z)}{2} \left[  \xi ^2\left(   \ln\frac{\tilde{\mu} _c^2}{\zeta _u^2 }   -2 \ln \frac{\tilde{\mu} _c^2\xi}{k_{\perp }^2} 
\right)
+ \gamma(\xi) \left(  -\frac{1}{2} \ln \left(\tilde{b}^2 \tilde{\mu} _c^2\right) \ln \frac{\tilde{\mu}
   _c^2}{\tilde{b}^2 \zeta _u^4  }     
    + \ln \frac{\tilde{\mu} _c^2}{k_{\perp }^2} \ln
  \frac{\tilde{\mu} _c^2 \xi^2}{k_{\perp }^2}
     \left. \left.
  +\ln ^2\xi +\frac{\pi ^2}{4} 
  \right) \right]
   \right. \right.
 \nonumber \\ &&
 \left.
+ \gamma(\xi) \left[
  -  \left( \frac{1}{z}\right)_+ \left(   \ln \tilde{b}^2 k_{\perp }^2  +\ln \frac{x}{\xi  \bar{\xi }^2} \right) 
  -  \left( \frac{\ln z}{z}\right)_+ 
  +  \frac{1}{\bar{\xi}}    \left(  \ln \tilde{b}^2 k_{\perp }^2   +\ln \frac{x z}{\xi  \bar{\xi }^2}  \right)
\right]
 \right\} + {\cal O}(b), \nonumber\\
 {\mathcal F}_{q/q}\Big\vert_{2b} &= &\frac{e^2  \alpha _s C_F }{\pi  k_{\perp }^2} \left\{ 
  \frac{4 \gamma(\xi) }{\epsilon _c^2} \delta (z)
  + \left(\frac{2}{\epsilon _c}+ \ln \frac{\tilde{\mu} _c^2 \xi }{k_{\perp }^2}   \right)\left[ \delta (z) \left(  \gamma(\xi) \ln \frac{\tilde{\mu} _c^2 \xi }{k_{\perp }^2} -\xi^2 \right)
+  \frac{1 }{\bar{z}}\left(- \frac{\gamma(\xi)}{(z)_+} +x \bar{\xi }+x+2\right)\right]
\right.
\nonumber \\
&&+\frac{1}{\bar{z}}
 \biggr [
\left(- \frac{\gamma(\xi)}{(z)_+} +x \bar{\xi } +x+2 \right)   \ln \frac{  \bar{\xi }}{z \bar{z}}
+  \left(\frac{1}{z} \right)_+ \left(\xi ^2- \gamma(\xi) \ln z\right)+\gamma(\xi) \left(\frac{\ln z}{z} \right)_+  -\xi ^2\biggr ]
\nonumber \\&&
\left .  + \frac{\gamma(\xi)  \delta(z) }{2}
\left( -\ln^2 \frac{\tilde{\mu} _c^2 \xi}{k_{\perp }^2}  
+\frac{ \pi^2}{6} \right)
\right\} + {\cal O}(b), \nonumber\\
{\mathcal F}_{q/q}\Big\vert_{2c}  &= &\frac{e^2  \alpha _s C_F }{\pi  k_{\bot }^2 } 
  \left[        
  \frac{\xi  x}{\bar{z}} \left( \frac{2}{\epsilon _c} + \ln \frac{\tilde{\mu} _c^2}{k_{\bot }^2} \right)
  + \left(\frac{\xi  x}{\bar{z}}+\frac{\bar{x}}{2}\right)   \ln \frac{\xi  x}{z \bar{z}^2}
  +\frac{\bar{x}}{2}  \ln \frac{x z}{\xi  \bar{\xi }^2}
  +\frac{\gamma(\xi)  x+\xi }{\bar{\xi } \bar{z}}
  -\frac{\bar{z}}{\bar{\xi }}   +\xi -x
    \right] 
  + {\mathcal O}(b), 
\nonumber\\    
   {\mathcal F}_{q/q}\Big\vert_{2d} &= &\frac{e^2  \alpha _s C_F }{\pi  k_{\bot }^2 
  \bar{\xi}^2} \left[  -\gamma(\xi)  z \left(  \ln \tilde{b}^2 k_\perp^2 +\ln \frac{x z}{\xi\bar{\xi}^2}+1\right) +\xi x\right]+ {\mathcal O}(b^2), 
\nonumber\\  
 {\mathcal F}_{q/q} \Big\vert_{2e} &= &\frac{e^2  \alpha _s C_F }{\pi  k_{\bot }^2 } 
  \left[        
  -z\left( 1+\frac{x^2}{\bar{z}^2}\right)  \left(    \frac{2}{\epsilon _c} +  \ln \frac{\tilde{\mu} _c^2}{k_{\bot }^2 } +\ln \frac{\xi  x}{z \bar{z}^2} \right)
  +\frac{\xi  \bar{\xi }}{\bar{z}}
  \right]   + {\mathcal O}(b), 
\nonumber\\   
  {\mathcal F}_{q/q}\Big\vert_{2f} &= &\frac{e^2  \alpha _s C_F }{\pi  k_{\bot }^2} \delta (z)  \left(   \frac{2}{ \epsilon_c} \gamma(\xi)    + \gamma(\xi)  \ln (\tilde{b}^2 \tilde{\mu}^2_c ) - \xi^2 \right) + {\mathcal O}(b), 
\end{eqnarray} 
with the $+$-distribution defined as in Eq.(\ref{NPLUS}). In both real- and virtual corrections there are divergent contributions given by terms containing  poles in $\epsilon_c$.  In our results, the U.V. divergences 
are subtracted or regularized by the nonzero impact parameter $b$. The light-cone divergences 
are regularized by non-light cone gauge links. In the limit $d\to 4$ we have only the divergent contributions 
which have poles in $\epsilon_c$.   
When we  sum these divergent contributions, we find that 
all double poles are cancelled and the sum contains only collinear single poles: 
\begin{eqnarray} 
{\mathcal F}_{q/q}  \biggr\vert_{div.}   = \frac{ e^2 \alpha_s C_F}{\pi}  \frac{1}{k_\perp^2} \biggr ( \frac{2}{\epsilon_c} \biggr )   \gamma (\xi/y)  \biggr (  
  -  \frac{3}{2} \delta (1-y ) - \frac{1+ y^2 }{(1- y)_+ }  \biggr ),    
    \quad y=x+\xi
\end{eqnarray}
with the $+$-distribution as the standard one given in Eq.(\ref{STP}). 
Using the result of the quark distribution of a quark as the target  in Eq.(\ref{qPDF}), 
we find that the collinear divergent part of the one-loop contribution of ${\mathcal F}_{q/q} $ from Fig.2 and Fig.3 can be factorized into the quark distribution function. Therefore, the one-loop contribution to the perturbative coefficient 
function $\hat {\mathcal H}_q$ is finite.  We find that our one-loop result of $\hat{{\cal H}}_q$ 
can be written in the form:  
 \begin{eqnarray}
\hat{{\cal H}}_q^{(1)}(x,b, \xi, k_\perp)&=& - \frac{e^2  \alpha _s C_F }{\pi  k_{\perp }^2} \gamma(\xi) 
\biggr  [\gamma(\bar y/\bar \xi)\left( \frac{1}{(\bar y)_+} -\delta(\bar y)\ln\bar \xi\right) \ln ( \tilde b^2 \mu^2) 
+ \frac{1}{2} \delta(\bar y) \biggr  (\ln ^2 (\tilde b^2 \zeta_u^2  e^{-1})
\nonumber\\ 
    &&  +3\pi^2+3\biggr  )\biggr  ]  +{{\cal H}}_q^{(1)}(x,\xi, k_\perp)    + {\cal O}(\alpha_s^2), 
 \end{eqnarray}
where ${{\cal H}}_q^{(1)}$ is given in Sect. 3.  
\par 
At the order of $\alpha_s$, the perturbative coefficient functions $\hat {\mathcal H}_g$ and $\hat {\mathcal H}_{gq}$ 
become nonzero. They are determined by calculating the fracture function of a gluon. The contribution is represented by diagrams in Fig.4. 
It is noted that there is no U.V. divergence and $\zeta_u$-dependence in the contribution. Because of that 
it is U.V. finite, 
the contribution from each diagram in Fig.4 in the small-$b$ limit is the same as given in Eq.(\ref{IFIG3}). 
Therefore, the factorization is the same as discussed in Sect.3.  Because of this,  the perturbative coefficient functions $\hat {\mathcal H}_g$ and  $\hat {\mathcal H}_{gq}$ 
are the same as ${\mathcal H}_g$ and  ${\mathcal H}_{gq}$ given in the last section, respectively.
 We have: 
\begin{equation} 
\hat {\mathcal H}_g ={\mathcal H}_g + {\mathcal O}(\alpha_s^2), \quad  \hat {\mathcal H}_{gq} = {\mathcal H}_{gq} + {\mathcal O}(\alpha_s^2). 
\end{equation}
Beyond one-loop level, they become different. 

\par 
The factorization given in Eq.(\ref{FACFFB}) is for the case for $k_\perp\gg \Lambda_{QCD}$. The $k_\perp$-behavior is determined by the perturbative coefficient functions given in the above.  If $k_\perp$ 
is at the order of $\Lambda_{QCD}$,  the behavior can not be predicted by perturbative QCD.  
In this case, another factorization is needed. 
By comparing the result of the integrated fracture function, we find that the TMD fracture function 
can be  factorized with the integrated fracture function as: 
\begin{equation}
{\mathcal F}_{q/h_A} (x, b, \xi, k_\perp) = \int_x^{1-\xi} \frac{dy}{y} C_q(x/y, b,	  \zeta_u) F_{q/h_A}  (y,\xi, k_\perp), 
\end{equation} 
where $C_q$ is the perturbative coefficient function. It is given by:
\begin{equation} 
C_q(x,b,\zeta_u) =  \delta(1-x)  -\frac{\alpha_s C_F }{4\pi} 
\left[\frac{2(1+x^2)}{(1-x)_+}\ln ( \tilde b^2 \mu^2) +\delta(1-x) \left(\ln ^2 (\tilde b^2 \zeta_u^2 e^{-1})+3\pi^2+3\right)\right ]   + {\mathcal O}(\alpha_s^2). 
\end{equation}
This formula is for the case of $k_\perp\sim\Lambda_{QCD}$. The $k_\perp$-behavior is described by 
the integrated fracture function which is nonperturbative.

\par 

From our results represented in this section, we can directly derive the evolution equations of $\mu$ and $\zeta_u$. 
The $\mu$-evolution is given by: 
\begin{eqnarray} 
 \frac{\partial {\mathcal F}_{q/h_A}  (x, b, \xi,  k_\perp, \zeta_u, \mu ) }{\partial \ln \mu} = 2 \gamma_F  {\mathcal F}_{q/h_A}  (x,  b, \xi,  k_\perp, \zeta_u, \mu ),    \quad  \gamma_F = \frac{ 3 \alpha_s C_F}{4 \pi} +{\mathcal O}  (\alpha_s^2),    
\end{eqnarray} 
where $\gamma_F$ is the anomalous dimension of quark fields in axial gauge. 
The evolution equation of $\zeta_u$ is called as Collins-Soper equation\cite{CSS}, which is very useful for resummation of large log's in perturbation theory. 
 From our explicit result 
we obtain Collins-Soper equation of the fracture function: 
\begin{eqnarray} 
 \frac{\partial {\mathcal F}_{q/h_A}  (x,  b, \xi,  k_\perp, \zeta_u, \mu ) }{\partial \ln\zeta_u }     &=& -\frac{\alpha_sC_F  }{ \pi}  \biggr ( \ln \frac{  \zeta_u^2 b^2 e^{2{\gamma} -1}}{4} \biggr ) {\mathcal F}_{q/h_A}  (x, b, \xi,  k_\perp, \zeta_u, \mu ) + {\mathcal O}(\alpha_s^2).    
\end{eqnarray} 
These two evolution equations are exactly the same as those of quark TMD parton distributions, as 
expected.  It is noted that one can define subtracted fracture function with the soft factor as discussed 
in Sect.2. for TMD quark distributions. In this case, the Collins-Soper equation can be derived 
only from the soft factor, whose results at three-loop are derived in \cite{YLHZ, AAV}. 

\par

\par\vskip20pt
\noindent 
{\bf 5. Summary} 
\par 
We have discussed factorizations of production of a lepton pair combined with a diffractively produced photon in hadron-hadron collisions.  In different kinematic regions factorizations can be made with different fracture functions. 
We take the diffractively produced particle as a photon to show at one-loop level that TMD- and integrated  quark fracture function can be factorized with standard parton distribution functions and fragmentation function, if the transverse momentum of the produced photon is much larger than $\Lambda_{QCD}$. 
In the $b$-space with the small transverse momentum of the photon, the TMD fracture function is factorized with the integrated fracture function. From our explicit calculations, the renormalization group equation of integrated fracture function and Collins-Soper equation of TMD fracture function are derived.
The derived equations are in agreement with expected.  Our main results provide a connection between 
factorizations with fracture functions and those with twist-2 parton distribution functions and fragmentation 
functions. They will be helpful for resummations of large log terms in collinear factorizations of relevant processes and building phenomenological models of fracture functions.

\vskip20pt
\noindent
{\bf Acknowledgments}
\par
The work is supported by National Natural
Science Foundation of P.R. China(No.11675241,11821505). The work of  K.B. Chen is supported by China Postdoctoral Science Foundation(No.2018M631588). The partial support from the CAS center for excellence in particle 
physics(CCEPP) is acknowledged.

\par\vskip40pt
\par

\end{document}